\documentclass[PRL,twocolumn,superscriptaddress,showpacs,preprintnumbers,amsmath,amssymb,floats]{revtex4}
\usepackage{amssymb}
\usepackage{graphicx}

\begin{document}

\title{Heavily electron-doped electronic structure and isotropic superconducting gap in A$_{x}$Fe$_{2}$Se$_{2}$ (A=K,Cs)}

\author{Y. Zhang}\author{L. X. Yang}\author{M. Xu}\author{Z. R. Ye}\author{F. Chen}\author{C. He}\author{J. Jiang}\author{B. P. Xie}
\affiliation{State Key Laboratory of Surface Physics,  Key Laboratory of Micro
and Nano Photonic Structures (MOE), and Department of Physics, Fudan
University, Shanghai 200433, People's Republic of China}

\author{J. J. Ying}\author{X. F. Wang}\author{X. H. Chen}

\affiliation{Hefei National Laboratory for Physical Sciences at Microscale and Department of Physics, University of Science
and Technology of China, Hefei, Anhui 230026, People¡¯s Republic of China}

\author{J. P. Hu}
\affiliation{Department of Physics, Purdue University, West Lafayette, Indiana
47907, USA}

\author{D. L. Feng}\email{dlfeng@fudan.edu.cn}
\affiliation{State Key Laboratory of Surface Physics,  Key Laboratory of Micro
and Nano Photonic Structures (MOE), and Department of Physics, Fudan
University, Shanghai 200433, People's Republic of China}

\begin{abstract}
The low energy band structure and Fermi surface of the newly
discovered superconductor, A$_{x}$Fe$_{2}$Se$_{2}$ (A=K,Cs), have
been studied by angle-resolved photoemission spectroscopy. Compared
with iron pnictide superconductors, A$_{x}$Fe$_{2}$Se$_{2}$ (A=K,Cs)
is the most heavily electron-doped with $T_c\sim$30~K. Only electron
pockets are observed with an almost isotropic superconducting gap of
$\sim$ 10.3~meV, while there is no hole Fermi surface near the zone
center, which indicates the inter-pocket hopping or Fermi surface
nesting is not a necessary ingredient for the unconventional
superconductivity in iron-based superconductors. Thus, the  sign
changed s$_\pm$ pairing  symmetry, a leading candidate proposed for
iron-based superconductors,   becomes conceptually irrelevant in
describing the superconducting state here. A more conventional
s-wave  pairing is a better  description.
\end{abstract}


\maketitle


\begin{figure}
\centerline{\includegraphics[width=6cm]{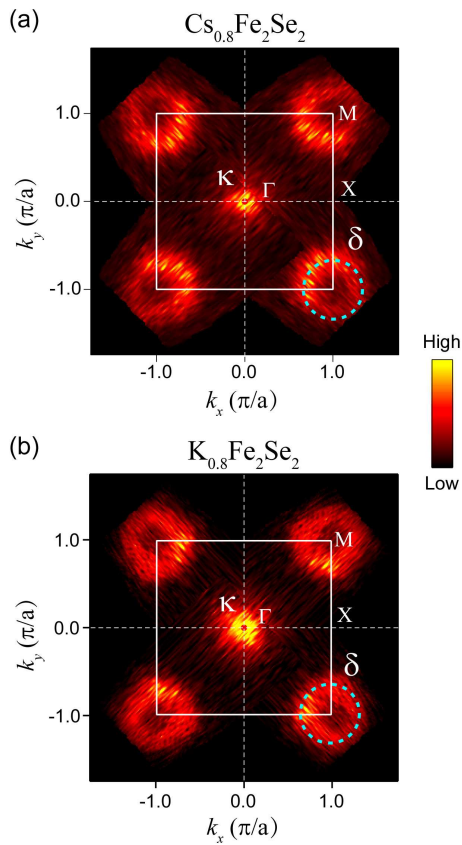}} \caption{The
four-fold symmetrized photoemission intensity map at the Fermi
energy ($E_F$) for (a) Cs$_{0.8}$Fe$_{2}$Se$_{2}$, and (b)
K$_{0.8}$Fe$_{2}$Se$_{2}$. The intensity was integrated over a
window of [$E_F$-15~meV, $E_F$+15~meV].} \label{fig.1}
\end{figure}

Since the discovery of iron-based high temperature superconductors
in 2008 \cite{Kamihara,Chen_Sm1},  the pairing mechanism of this new
class of materials has been extensively studied \cite{mazin, seo,
wang, Kuroki,HDing,BaCo,YZhangBK}. Among various aspects that were
suggested to be essential for the unconventional superconductivity,
the multi-band nature of the electronic structure arguably received
the most attention  \cite{YZhang_orbital}. In particular, in the
weak coupling approach, it was suggested that the superconductivity
could be boosted by the nesting or hopping between the electron-like
Fermi surfaces near the zone corner and the hole-like Fermi surfaces
near the zone center \cite{BaCo}, and the sign change of the
superconducting order parameters between the hole and electron Fermi
surfaces stems from the inter-pocket scattering \cite{mazin}.
Moreover,  The gap symmetry and gap anisotropy (e.g. the appearance
of nodes) are sensitive to the presence of certain Fermi surface and
the interactions among the bands \cite{theory}. However, in the
strong coupling approach, the superconducting pairing is dominated
by the intra-orbital pairing and relatively robust against the
change of band structures \cite{seo}. The dominating pairing symmetry
is an $A_{1g}$ s-wave. Experimentally, both nodal and nodeless gap
distributions have been reported by different techniques in various
systems. Although the scanning tunneling spectroscopy (STS) data on
Fe(Se,Te) suggest a nodeless $s_{\pm}$-wave gap \cite{Hanaguri}, the
STS data on FeSe indicate a nodal gap \cite{QKXue}. While
photoemission data on Ba$_{1-x}$K$_x$Fe$_2$As$_2$ and
BaFe$_{2-x}$Co$_x$As$_2$ fit the $s_{\pm}$-wave gap function well
\cite{HDing,BaCo,YZhangBK}, the thermal conductivity measurements
suggest gap in heavily hole doped KFe$_2$As$_2$ to be a nodal type
\cite{KFe2As2}. A conclusion on the pairing behaviors seems to be still far-fetched for the iron based superconductors.

Recently, a new series of iron-based superconductors,
A$_{x}$Fe$_{2}$Se$_{2}$ (A=K,~Cs),  has been discovered with
relatively high transition temperature of $\sim$30~K
\cite{ChenXL,Yoshi,Krz,JJYing}. Judging from their chemical formula,
these compounds would be the most heavily electron-doped amongst the
iron-based superconductors. More importantly, it provides an
opportunity to examine the common aspects of the electronic
structure and pairing mechanism in iron-based systems from the most
electron-doped end. In this paper, we report the angle resolved
photoemission spectroscopy (ARPES) study of  A$_{x}$Fe$_{2}$Se$_{2}$
(A=K,~Cs). We found that they are indeed the most heavily
electron-doped iron-based superconductor. Only electron pockets are
observed, while there is no hole Fermi surface near the zone center.
The superconducting gap around the electron pockets is isotropic,
and about 10.3~meV, i.e. $\sim$~4$k_B T_c$ for both compounds. Our
result indicates the inter-pocket hopping or Fermi surface nesting
is not a necessary ingredient for the unconventional
superconductivity in iron-based superconductors. Thus the sign
change in $s_{\pm}$-wave pairing, a promising pairing symmetry
candidate suggested earlier, is not a fundamental property of
iron-based superconductors. Rather, the isotropic gap structure is
better to be considered as a more conventional s-wave.


A$_{0.8}$Fe$_{2}$Se$_{2}$ (A=K,~Cs, nominal composition) single
crystals were synthesized by self-flux method as described elsewhere
in detail \cite{JJYing}, which show flat shiny surfaces with dark
black color. Fitting the X-ray diffraction data assuming the I4/mmm
symmetry gave $a = 3.8912$~\AA, $c = 14.1390$~\AA~~for
K$_{0.8}$Fe$_{2}$Se$_{2}$, and $a = 3.9618$~\AA, $c =
15.285$~\AA~~for Cs$_{0.8}$Fe$_{2}$Se$_{2}$.
K$_{0.8}$Fe$_{2}$Se$_{2}$ shows  the onset superconducting
transition temperature ($T_c$) of 31.7~K, and it reaches zero
resistivity at 31.2~K; while Cs$_{0.8}$Fe$_{2}$Se$_{2}$  shows the
onset $T_c$ of 30~K with a transition width of 2~K in the
resistivity data.   The magnetic susceptibility measurements showed
almost 100\% shielding fraction, indicating the bulk superconducting
nature and good quality of the crystals.  ARPES measurements were
performed with SPECS UVLS discharge lamp (21.2~eV He-I$\alpha$
light) and a Scienta R4000 electron analyzer. The overall energy
resolution was set to 9 or 12~meV, and the angular resolution was
0.3 $^{\circ}$. The sample was cleaved \textit{in situ}, and
measured under ultra-high-vacuum of $5\times10^{-11}$\textit{torr}.
The actual chemical compositions were determined by energy
dispersive X-ray (EDX) spectroscopy  on the cleaved surface after
the photoemission measurements,  which gave
K~:~Fe~:~Se~=~0.94~:~1.98~:~2, and Cs~:~Fe~:~Se~=~0.92~:~1.99~:~2.
For simplicity, they are called by the nominal compositions
throughout the text.


\begin{figure}
\centerline{\includegraphics[width=8.5cm]{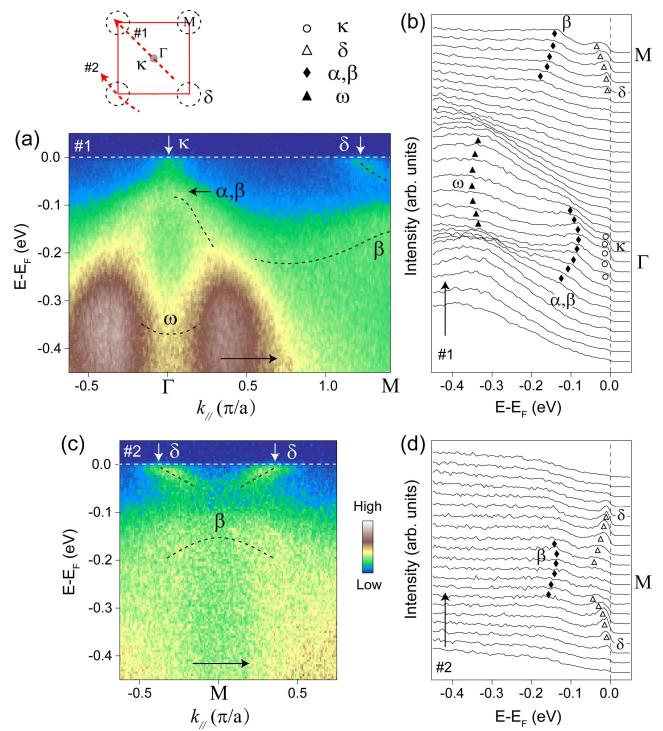}}
\caption{Photoemission data of K$_{0.8}$Fe$_{2}$Se$_{2}$. (a) The
photoemission intensity, and (b) energy distribution curves (EDC's)
along cut~\#1 or $\Gamma$-M   in the projected two dimensional
Brillouin zone as shown at the top of panel a. (c) and (d) The
photoemission intensity, and EDC's along cut~\#2 across the zone
corner respectively.} \label{fig.2}
\end{figure}

Figure~\ref{fig.1} shows the photoemission intensity maps at the
Fermi energy ($E_F$) for Cs$_{0.8}$Fe$_{2}$Se$_{2}$ and
K$_{0.8}$Fe$_{2}$Se$_{2}$. The Fermi surface topology is similar in
both systems. There is an electron-like Fermi pocket surrounding the
zone corner, and some spectral weight is located at the zone center.

The sizes of the electron pocket are about the same for both K and
Cs compounds, and their band structures  are very much alike as
well. Figure~\ref{fig.2} further reveals the  band structure of
K$_{0.8}$Fe$_{2}$Se$_{2}$. Around the zone center
[Figs.~\ref{fig.2}(a) and ~\ref{fig.2}(b)], there is a small
electron-like feature, the $\kappa$ band. The spectral weight of
this band is rather weak, indicating that it might be a tail of
certain band slightly above $E_F$.  This is consistent with a recent
band structure calculation that shows an electron pocket in the zone
center for K$_{x}$Fe$_{2}$Se$_{2}$ at certain doping \cite{LDA}.  At
-0.1~eV and below, the fast dispersive features are most likely the
$\alpha$ and $\beta$ bands that form the hole-like pockets in
Fe(Te,Se) \cite{ChenFeiPRB} and iron pnictides. Moreover, the
$\omega$ band is observed around 0.35~eV below $E_F$ near $\Gamma$,
and similar band is  observed in iron pnictides and known to be made
of the $d_{z^2}$ orbitals \cite{YZhang_orbital}.  Around the zone
corner [Figs.~\ref{fig.2}(c) and \ref{fig.2}(d)], an electron-like
band, $\delta$, is observed together with the $\beta$ band from the
zone center.  Theoretically, two electron-like bands around M were
predicted for K$_{0.8}$Fe$_{2}$Se$_{2}$ \cite{LDA}, just like for
the iron pnictides. However, at certain experimental geometry, only
one electron-like band is often observed due to the matrix element
of the $3d$ orbitals \cite{YZhang_orbital}. It is also predicted
that the Fermi crossings of these two bands are almost-degenerate,
and the bottom of the other electron like band would coincide with
the $\beta$ band at M \cite{LDA, YZhang_orbital}.  In general, the
bands around $\Gamma$ resemble those observed in Fe(Te,Se)  and iron
pnictides, except that the chemical potential is shifted up via
electron doping here.  Take BaFe$_{1.85}$Co$_{0.15}$As$_2$ for a
comparison, although it is optimally electron-doped with a $T_c$ of
about 25~K,  there are still several hole-like pockets around the
zone center \cite{YZhang_orbital}. Moreover, the binding energy of
the $\omega$ band near $\Gamma$ is about 200~meV larger in
K$_{0.8}$Fe$_{2}$Se$_{2}$ than that  in
BaFe$_{1.85}$Co$_{0.15}$As$_2$  due to the higher electron doping in
the former compound. However, it is interesting to note that the
$\delta$ band around M is rather flat and shallow in
K$_{0.8}$Fe$_{2}$Se$_{2}$, indicative of a non-rigid band behavior.

Assuming the electronic structure of A$_{0.8}$Fe$_{2}$Se$_{2}$ to be
two dimensional and two almost-degenerate electron pockets around M,
in the ionic picture, we could obtain $3d^{6.18}$ per Fe ion for
A$_{0.8}$Fe$_{2}$Se$_{2}$ by the measured Fermi surface volume. This
cannot account for all the electrons in the $3d^{6.45}$
configuration calculated directly from their actual chemical
compositions. Therefore, the inconsistency may suggest possible
$k_z$ dispersions of the $\delta$ and $\kappa$ bands.  To fully
reveal the Fermi surface topology, more detailed $k_z$ and
polarization dependence studies are required with variable photon
energies at a synchrotron facility.  In any case, the measured
electronic structure clearly shows that A$_{0.8}$Fe$_{2}$Se$_{2}$ is
indeed the most heavily electron-doped iron-based superconductor by
far.

\begin{figure}
\centerline{\includegraphics[width=8cm]{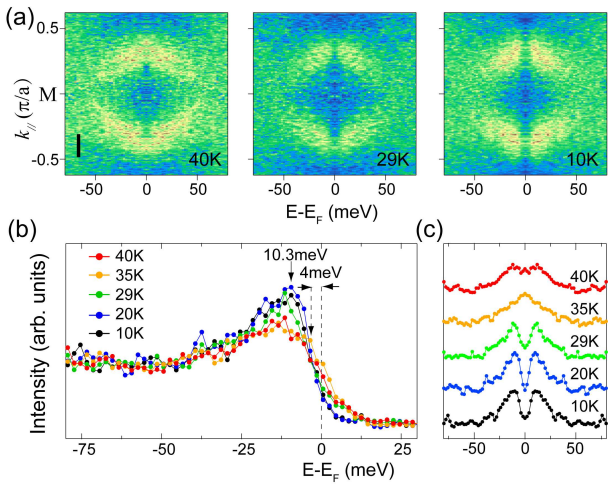}} \caption{(a) The
symmetrized (with respect to  $E_F$) photoemission intensity of
K$_{0.8}$Fe$_{2}$Se$_{2}$ at 40, 29, and 10~K along the cut \#2 in
Fig.~\ref{fig.2}. (b) Temperature dependence of the spectrum
integrated over the  momentum region indicated by the  black bar in
panel a, and (c) its stacked symmetrized version. Energy resolution
was set to 9~meV in the gap measurements.} \label{SC}
\end{figure}

\begin{figure}
\centerline{\includegraphics[width=8cm]{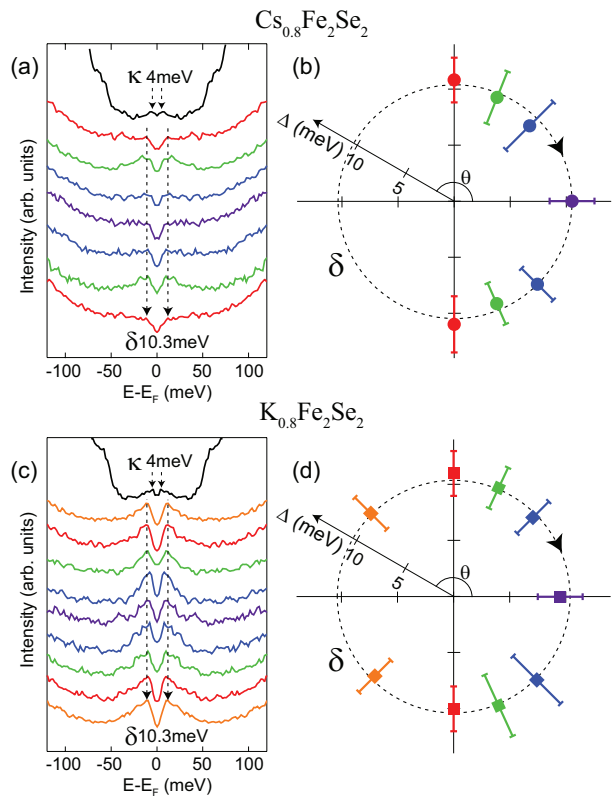}} \caption{(a) The
symmetrized (with respect to $E_F$) EDC's at various Fermi crossings
for Cs$_{0.8}$Fe$_{2}$Se$_{2}$. The top curve is for the $\kappa$
band at $\Gamma$, and the rest are for the $\delta$ band with
momenta clockwise along the $\delta$ pocket as shown by the arrows
in panel b. (b) Gap distribution of the $\delta$ band around M in a
polar coordinate,  where the radius represents the gap, and the
polar angle $\theta$ represents the position on the $\delta$ pocket
with respect to M, with $\theta=0$ being the M-$\Gamma$ direction.
(c) and (d) are the same as panels a and b except for
K$_{0.8}$Fe$_{2}$Se$_{2}$. All data were taken at 9K.} \label{gap}
\end{figure}

To study the superconducting gap of K$_{0.8}$Fe$_{2}$Se$_{2}$, high
resolution data  taken above and below $T_c$ along the cut~\#2 are
compared in Fig.~\ref{SC}. As shown in Fig.~\ref{SC}(a), there is no
gap at $E_F$ near the Fermi momentum in the normal state, while a
clear gap shows up at low temperatures in the superconducting state.
The temperature dependence of the spectrum illustrates that the
spectral weight near the Fermi energy is depleted and a coherent
peak feature grows with decreasing temperature [Fig.~\ref{SC}(b)].
One can clearly identify a leading edge gap of 4~meV at 10~K. Taking
the coherent peak position  as the superconducting gap,  we find it
to be $\sim10.3$~meV, \textit{i.e.}  $\sim4 k_B T_c$. This ratio
between gap and $T_c$ falls into the same regime as other iron-based
superconductors \cite{HDing,BaCo,YZhangBK}.  The symmetrized spectra
in  Fig.~\ref{SC}(c) further indicate that the gap disappears when
the temperature is above $T_c$.  We note that the spectra are
integrated over a small region around the normal state Fermi
momentum to compensate the relatively weak signal, but the estimated
gap amplitude would not be affected much, as the gap feature in this
region is rather flat [Fig.~\ref{SC}(a)].

By examining the symmetrized EDC's in the superconducting state at
various  Fermi crossings of both the $\kappa$ and $\delta$ bands,
the momentum distribution of superconducting gap is deduced in
Fig.~\ref{gap}. For both K and Cs compounds,  the gap of the
$\delta$ band around the M point is of the isotropic $s$-wave type
within the experimental uncertainty, which averagely is about
10.3~meV; while the gap is 4~meV  for the $\kappa$ band at $\Gamma$.
The smaller gap at $\Gamma$ than those around M certainly violates
the simple gap function of $\cos k_x \cos k_y$ for $s_{\pm}$-wave
order parameter, and indicates the gaps may  be orbital dependent.
Moreover, since the spectral weight near the zone center is minimal,
its contribution to the superconductivity would be rather negligible
with such a small gap. Therefore, the inter-pocket scattering
previously suggested can not be the essential force driving the
superconductivity.

In summary, our data show that the rather robust superconductivity
in such a highly  electron-doped iron-based superconductor could
mainly rely on the electron Fermi surfaces near M. The  rather
unique electronic structure in  A$_{0.8}$Fe$_{2}$Se$_{2}$ (A=K,~Cs)
further highlights the diversity of the iron-based superconductors,
and suggests that the superconductivity is very robust against the
change of Fermi surfaces. Our data also strongly suggest that the
inter-band hopping might not be so substantial as previous data
suggested. Thus, the promising candidate, the so called s$_\pm$ wave
characterized by the sign change of the superconducting orders
between electron and hole pockets, is not a proper description of
the superconducting state in A$_{x}$Fe$_{2}$Se$_{2}$. Instead,  the
more conventional s-wave type is a more proper and general
description for the iron-based superconductors. Our result can also
be viewed as a support to the picture derived from the strong
coupling approach suggesting the pairing being the intra-orbital
pairing caused by local electron-electron correlation.



This work was supported by the NSFC, MOE, STCSM, and National Basic Research
Program of China (973 Program)  under the grant No. 2011CB921802.


\begin{thebibliography}{10}

\bibitem{Kamihara} Y. Kamihara, T. Watanabe, M. Hirano, and H. Hosono, J. Am. Chem. Soc. \textbf{130}, 3296 (2008).

\bibitem{Chen_Sm1} X. H. Chen, T. Wu, G. Wu, R. H. Liu, H. Chen, and D. F. Fang, Nature (London) \textbf{453}, 761 (2008).

\bibitem{mazin} I.~I. Mazin, {\it et ~al}, Phys. Rev. Lett. {\bf101},  057003 (2008).

\bibitem{seo} K. Seo, B.~A. Bernevig, and J. Hu, Phys. Rev. Lett. {\bf 101},  206404  (2008).

\bibitem{wang} F. Wang {\it et~al.}, Phys. Rev. Lett. {\bf 102},  1047005  (2009).


\bibitem{Kuroki}K. Kuroki, S. Onari, R. Arita, H. Usui, Y. Tanaka, H. Kontani, and H. Aoki, Phys. Rev. Lett. \textbf{101}, 087004 (2008).

\bibitem{HDing} H. Ding, P. Richard, K. Nakayama, K. Sugawara, T. Arakane, Y. Sekiba, A. Takayama, S. Souma, T. Sato, T. Takahashi, Z. Wang, X. Dai, Z. Fang, G. F. Chen, J. L. Luo, and N. L. Wang, Europhys. Lett. \textbf{83}, 47001 (2008).

\bibitem{BaCo} K. Terashima, Y. Sekiba, J. H. Bowen, K. Nakayama, T. Kawahara, T. Sato, P. Richard, Y.-M. Xu, L. J. Li, G. H. Cao, Z.-A. Xu, H. Ding and T. Takahashi, Proc. Natl. Acad. Sci. U.S.A. \textbf{106}, 7330
(2009).

\bibitem{YZhangBK}Y. Zhang, L. X. Yang, F. Chen, B. Zhou, X. F. Wang, X. H. Chen, M. Arita, K. Shimada, H. Namatame, M. Taniguchi, J. P. Hu, B. P. Xie and D. L. Feng, Phys. Rev. Lett. \textbf{105}, 117003 (2010).

\bibitem{YZhang_orbital} Y. Zhang, B. Zhou, F. Chen, J. Wei, M. Xu, L. X. Yang, C. Fang, W. F. Tsai, G. H. Cao, Z. A. Xu, M. Arita, H. Hayashi, J. Jiang, H. Iwasawa, C. H. Hong, K. Shimada, H. Namatame, M. Taniguchi, J. P. Hu, D. L. Feng,  Phys. Rev. {\bf B} (in press), preprint availabe as arXiv:0904.4022.

\bibitem{theory} V. Vildosola, L. Pourovskii, R. Arita, S. Biermann, and A. Georges, Phys. Rev. B 78, 064518 (2008); K. Kuroki, arXiv:1008.2286v1; R. Thomale et al., Phys. Rev. B 80, 180505 (2009); R. Thomale, C. Platt, W. Hanke, and B. A. Bernevig, arXiv:1002.3599v1

\bibitem{Hanaguri} T. Hanaguri, S. Niitaka, K. Kuroki, and H. Takagi, Science \textbf{23}, 441-443 (2010).

\bibitem{QKXue} V-shaped scanning tunneeling spectra has been observed in the superconducting state of $\alpha$-FeSe by Q. K. Xue and co-workers (private communication).

\bibitem{KFe2As2} J. K. Dong, S. Y. Zhou, T. Y. Guan, H. Zhang, Y. F. Dai, X. Qiu, X. F. Wang, Y. He, X. H. Chen, and  S. Y. Li, Phys. Rev. Lett. \textbf{104}, 087005 (2010).

\bibitem{ChenXL} J. Guo, S. Jin, G. Wang, S. Wang, K. Zhu, T. Zhou, M. He, and X. Chen, Phys. Rev. \textbf{B 82}, 180520 (2010).

\bibitem{Yoshi} Yoshikazu Mizuguchi, Hiroyuki Takeya, Yasuna Kawasaki, Toshinori Ozaki, Shunsuke Tsuda, Takahide Yamaguchi, and Yoshihiko Takano, arXiv:1012.4950 (unpublished).

\bibitem{Krz} A. Krzton-Maziopa, Z. Shermadini, E. Pomjakushina, V. Pomjakushin, M. Bendele, A. Amato, R. Khasanov, H. Luetkens, and K. Conder,  arXiv:1012.3637 (unpublished).

\bibitem{JJYing} J. J. Ying, X. F. Wang, X. G. Luo,  A. F. Wang, M. Zhang, Y. J. Yan, Z. J. Xiang, R. H. Liu, P. Cheng, G. J. Ye, and X. H. Chen,  arXiv:submit/0169761 (unpublished).

\bibitem{LDA} I. R. Shein,  A. L. Ivanovskii, arXiv:1012.5164, (unpublished).

\bibitem{ChenFeiPRB}Fei Chen, Bo Zhou, Yan Zhang, Jia Wei, Hong-Wei Ou, Jia-Feng Zhao, Cheng He, Qing-Qin Ge, Masashi Arita,
Kenya Shimada, Hirofumi Namatame, Masaki Taniguchi, Zhong-Yi Lu,
Jiangping Hu, Xiao-Yu Cui, and D. L. Feng, Phys. Rev. B \textbf{81},
014526 (2010).

\end{thebibliography}
\end{document}